\documentclass{pasj00}

\usepackage{threeparttable}
\usepackage{natbib}
\usepackage{pdflscape}
\usepackage{graphicx}

\begin{document}
\SetRunningHead{R. A. Burns et al.}{H$_{2}$O masers in IRAS 20143+3634, and the Galactic constants}

\title{VLBI Observations of H$_{2}$O Maser Annual Parallax and Proper Motion in IRAS 20143+3634: Reflection on the Galactic Constants}

\author{Ross A.   \textsc{Burns},\altaffilmark{1}
        Yoshiyuki \textsc{Yamaguchi},\altaffilmark{1}
        Toshihiro \textsc{Handa},\altaffilmark{1}
        Toshihiro \textsc{Omodaka},\altaffilmark{1}
        Takumi   \textsc{Nagayama},\altaffilmark{2}
        Akiharu   \textsc{Nakagawa},\altaffilmark{1}
        Masahiko   \textsc{Hayashi},\altaffilmark{3,5}
        Tatsuya   \textsc{Kamezaki},\altaffilmark{1}
        James O.   \textsc{Chibueze},\altaffilmark{3}
        Makoto   \textsc{Shizugami},\altaffilmark{2}
        and
        Makoto   \textsc{Nakano},\altaffilmark{4}
        }
\altaffiltext{1}{Graduate School of Science and Engineering, Kagoshima University,\\
                 1-21-35 K\^orimoto, Kagoshima, Kagoshima 890-0065, Japan}
\email{RossBurns88@MilkyWay.sci.Kagoshima-u.ac.jp}
\altaffiltext{2}{Mizusawa VLBI Observatory, National Astronomical Observatory of Japan, \\
                 2-12 Hoshi-ga-oka, Mizusawa-ku, Oshu, Iwate 023-0861, Japan}
\altaffiltext{3}{National Astronomical Observatory of Japan,
2-21-1 Osawa, Mitaka, Tokyo 181-8588, Japan}
\altaffiltext{4}{Faculty of Education and Welfare Science, Oita University, Oita 870-1192, Japan}
\altaffiltext{5}{School of Mathematical and Physical Science, The Graduate University for Advanced Studies (SOKENDAI), Hayama, Kanagawa 240-0193, Japan}


%


\maketitle

\begin{abstract}

\noindent We report the results of VLBI observations of H$_{2}$O masers in the IRAS 20143+3634 star forming region using VERA (VLBI Exploration of Radio Astronomy). By tracking masers for a period of over two years we measured a trigonometric parallax of $\pi = 0.367 \pm 0.037$ mas, corresponding to a source distance of $D = 2.72 ^{+0.31}_{-0.25}$ kpc and placing it in the Local spiral arm. Our trigonometric distance is just 60\% of the previous estimation based on radial velocity, significantly impacting the astrophysics of the source. 
We measured proper motions of $−2.99 \pm 0.16$ mas yr$^{-1}$ and $−4.37 \pm 0.43$ mas yr$^{-1}$ in R.A. and Decl. respectively, which were used to estimate the peculiar motion of the source as $(U_{s},V_{s},W_{s}) = (-0.9 \pm 2.9, -8.5 \pm 1.6, +8.0 \pm 4.3)$ km s$^{-1}$ for $R_0=8$ kpc and $\Theta_0=221$ km s$^{-1}$, and $(U_{s},V_{s},W_{s}) = (-1.0 \pm 2.9, -9.3 \pm 1.5, +8.0 \pm 4.3)$ km s$^{-1}$ for $R_0=8.5$ kpc and $\Theta_0=235$ km s$^{-1}$. IRAS 20143+3634 was found to be located near the tangent point in the Cygnus direction. Using our observations we derived the angular velocity of Galactic rotation of the local standard of rest (LSR), $\Omega_{0} = 27.3 \pm 1.6$ km s$^{-1}$ kpc$^{-1}$, which is consistent with previous values derived using VLBI astrometry of SFRs at the tangent points and Solar circle. It is higher than the value recommended by the IAU of $\Omega_{0} = 25.9$ km s$^{-1}$ kpc$^{-1}$ which was calculated using the Galactocentric distance of the Sun and circular velocity of the LSR. 

\end{abstract}



\section{Introduction} 

The Galactic circular rotation velocity evaluated at the position of the Sun, $\Theta_{0}$, and the Galactocentric distance to the Sun, $R_{0}$, are two of the fundamental parameters used in discussing the structure and kinematics of the Milky Way. These parameters are central to estimating the kinetic distance of Galactic sources, thus influencing estimations of their physical properties. Furthermore, these parameters affect the shape of the Galactic rotation curve which has long been an essential tool for evaluating the dynamical mass distribution in the Milky Way.

The International Astronomical Union (IAU) recommends values of $R_{0} = 8.5$ kpc and $\Theta_{0}=220$ km s$^{-1}$ for the Galactocentric distance of the Sun, and Galactic circular velocity of the LSR \citep{Kerr86}. However a growing number of astrometric observations utilising very long baseline interferometer (VLBI) suggest the need to revise these values (\citet{Reid09,Honma12} \emph{and references therein}), further stressing the importance of VLBI astrometry in our understanding of the Milky Way.
It is possible to estimate $\Theta_{0}$ and $R_{0}$, hereby referred to as the Galactic constants, by measuring the distance and motion of many sources in the disk of the Milky Way, with respect to the local standard of rest (LSR). 
Historically, they have been derived from global sinusoidal patterns shown in the proper motions and radial velocities of nearby stars \citep{Oort27}. However, this method is based on the assumption that the systematic motion of the Solar neighbourhood does not deviate from Galactic circular motion.
To circumvent this problem we must extend the sampling region to beyond the optical observable range, which is limited by interstellar extinction. VLBI astrometry is a viable approach since radio observations do not suffer from interstellar extinction.

There are two approaches to estimating the Galactic constants. One method uses a sample of sources widely distributed on the Galactic disk. With a kinematic model of the Galaxy, multi-parameter fitting determines the Galactic constants. 
Although this approach has the advantage of statistically reducing the random motion inherent in the source sample by increasing the number of sources, the reliability of this method is undermined by the high number of variable parameters that are simultaneously solved for. 
The other approach focuses on sources at special locations where the Galactic constants  can be derived with a lower dependancy on the model of Galactic kinematics. Two such locations are positions on the Solar circle and at the tangent point, defined as the position closest to the Galactic center on the assigned line of sight.
Astrometric observations of a source near these locations can give the ratio of the constants, $\Omega_0=\Theta_0/R_0$, with appropriate accuracy even if we do not know the exact location of the source \citep{Nagayama11a,Ando11}.

IRAS 20143+3634 is an intermediate to high mass star forming region (SFR), as shown in this paper. Furthermore, IRAS 20143+3634 resides near the tangent point in the Cygnus direction. A source at the tangent point, with negligibly small deviation from the Galactic circular motion, moves only along the line of sight to the Sun. Thus we can infer that any lateral proper motion observed on the sky reflects only the Galactic rotation of the LSR, which leads to a simplified estimation of $\Theta_{0}$, independent of the shape of the rotation curve. In this work, our aim was to use the astrometry of H$_{2}$O masers in IRAS 20143+3634 to measure its trigonometric distance and motion in the Galactic plane. Using these results, and those of other sources at the tangent points and Solar circle, we investigated the Galactic constants with fewer assumptions on Galactic structure and kinematics. This simplified approach may be considered more robust than the multi-parameter fitting method.
IRAS 20143+3634 is an infrared source listed in the IRAS point source catalogue \citep{IRAS88}, undergoing intermediate to high mass star formation. This source exhibits a highly compact core, seen in CS$(J=2-1)$ observations \citep{Ao04}, and wide velocity wings in $^{12}$CO($J=1-0$) indicative of the presence of outflows \citep{Yang02}.

Disentangling the proper and parallactic motions of masers requires precise astrometric observations. Such precision is available through VLBI maser monitoring using VLBI Exploration of Radio Astronomy (VERA) \citep{Koba03}. All astrometric observations used in this paper were obtained by positionally referencing 22 GHz H$_{2}$O masers with respect to a well defined quasar position, which gives the source position in six dimensional phase-space.
The first detection of H$_{2}$O maser emission in IRAS 20143+3634 was made by \citet{Sunada07}. 
In this paper we present the first VLBI observations of the annual parallax and proper motions of H$_{2}$O masers in IRAS 20143+3634. Using these measurements we measured the distance of IRAS 20143+3634 from the Sun and evaluated the Galactic constants.

This paper continues as follows: Observations and data reduction are discussed in \S2. Results are reported in \S3, including the methods used to evaluate the parallactic, internal and systemic motions of masers, and motion of the driving source. In \S4 we re-evaluate the physical parameters of IRAS20143+3634 from archive data, using our new distance estimate. We then discuss in detail the evaluation of $\Omega_{0}$, the ratio of the Galactic constants, for a variety of methods and sources. Finally, we make comparisons between the observationally determined values of $\Omega_{0}$, and that which is derived from the ratio of the IAU recommended Galactic constants. Conclusions made in this paper are summarised in \S5.



\section{Observations and data reduction}

We observed H$_{2}$O maser emission in IRAS 20143+3634 over a period of about two years with a total of 11 observation epochs using VERA. 
We adopted a rest frequency of 22.235080 GHz for the H$_{2}$O(6-5) maser line. The phase tracking center of the source was set to $(\alpha, \delta)_{\mathrm{J}2000.0}=
(20^{\mathrm{h}}16^{\mathrm{m}}13^{\mathrm{s}}.3617$,
+36$^{\circ}$43'33".920). Using the dual-beam system installed on VERA we simultaneously observed J2015+3710, which is listed in the VLBA Calibrator Source List \citep{VLBA2}, as a positional reference. The typical on-source integration time for each source was about 3 hours per epoch. The exact position of the reference source J2015+3710 was assumed to be 
$(\alpha, \delta)_{\mathrm{J}2000.0}=
(20^{\mathrm{h}}15^{\mathrm{m}}28^{\mathrm{s}}.729794$,
+37$^{\circ}$10'59".51480). A typical flux of $\sim$1~Jy was recorded for the reference source during the period of our observations. The angular separation of IRAS 20143+3634 and J2015+3710 was $0.^{\circ}48$.

Intermittent observations of the bright continuum sources NRAO530 and BL Lac were made every $\sim$1.5 hours to provide bandpass and group delay calibration.
An artificial noise signal was injected into both the target and positional reference source beams to continuously correct the instrumental phase difference \citep{Honma08a}.
The total duration of a typical observation epoch was about 7-8 hours.

\begin{table*}[t]

\caption{Summary of observations made with VERA.\label{table:1}}
\begin{center}
\null
\small
\begin{tabular}{cccccc}
\hline
Observation&&Modified&Synthesised beam&Detected\\
Epoch&Date&Julian Date&mas $\times$ mas, PA&spots\\ \hline
1&2008 Dec 11&54811&1.3$\times$0.7, $-$51$^{\circ}$&2\\
2&2009 Feb 10&54872&1.3$\times$0.8, $-$56$^{\circ}$&8\\
3&2009 May 29&54980&1.3$\times$0.7, $-$58$^{\circ}$&12\\
4&2009 Sep 26&55100&1.2$\times$0.9, $-$44$^{\circ}$&6\\
5&2009 Dec 14&55179&1.3$\times$0.8, $-$45$^{\circ}$&9\\
6&2010 May 26&55345&1.4$\times$0.8, $-$30$^{\circ}$&11\\
7&2010 Oct 17&55486&1.3$\times$0.9, $-$41$^{\circ}$&11\\
8&2010 Dec 26&55556&1.2$\times$0.7, $-$42$^{\circ}$&8\\
\hline
\end{tabular}

\end{center}

\end{table*}

Left-handed circular polarisation signals were sampled with 2-bit quantisation, and filtered with the VERA digital filter unit \citep{Iguchi05}. The 256 MHz total bandwidth of the recording system was divided into 16 intermediate frequency (IF) channels, with a bandwidth of 16 MHz each. One IF channel was allocated to the maser source while the remaining 15 IF channels were assigned to the reference source beam.

Interferometric correlation was processed using the Mitaka FX correlator
\citep{Chikada}. The frequency resolution for the H$_{2}$O maser line was set to 31.25 kHz,
corresponding to a velocity resolution of 0.42 km s$^{-1}$.
Since the a-priori delay model applied during correlation processing at the Mitaka FX correlator was not accurate enough for precise astrometry, the visibility phase was calibrated using a more accurate delay model based on recent results obtained in the analysis of geodynamics.
In this model, fluctuations in the visibility phase caused by the atmosphere are calibrated based on GPS measurements of the atmospheric zenith delay due to tropospheric water vapour
\citep{Honma08b}.

Data reduction was carried out using the Astronomical Image Processing System (AIPS) developed by the National Radio Astronomy Observatory (NRAO). In the primary stage of data reduction we performed a normalisation of the auto-correlated visibility data taken at each station using the AIPS task ACCOR.
Amplitudes were converted into system noise equivalent flux density in `Jy' using the gain and system temperatures recorded at each station using the task APCAL.
Doppler shifts of the maser emission due to the rotation of the Earth were corrected at each antenna station using CVEL. The fringe fitting of the reference source was performed by the task FRING. The positional reference source was self-calibrated by iterating the task pair IMAGR and CALIB. The final phase residual solutions were applied to the IRAS 20143+3634 data, appointing the positional reference source as the phase reference.

Using the calibrated data, maps were made for each velocity channel that showed emission by deconvolution of the modelled emission and beam pattern using the CLEAN procedure based on \citet{Hog74}. Finally the absolute peak position of each maser spot was determined using Gaussian fits applied to the CLEANed maser images, adopting a signal-to-noise ratio of 7 as our cut-off. Data of sufficient quality were collected on 8 epochs, the detections are summarised in Table~\ref{table:1} alongside the typical synthesised beam sizes. 


\section{Results}

\subsection{Distribution of H$_{2}$O masers in IRAS 20143+3634}
\begin{figure}[h!]
\begin{center}
\hspace*{-1.6cm}
\includegraphics[scale=0.9]{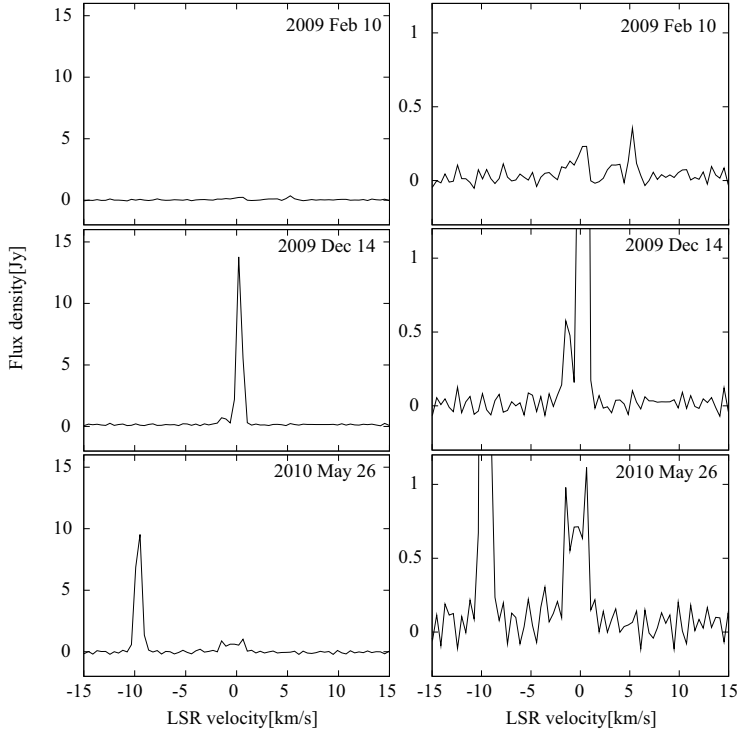}
\caption{Scalar-averaged cross-power spectra of IRAS 20143+3634 at three epochs as noted in the top right corner of each panel, for full scale of the flux density (\emph{left}), and expanded presentation (\emph{right}).\label{fig:1}}
\end{center}
\end{figure}

The scalar averaged cross-power spectra displaying the LSR velocity distribution of H$_{2}$O masers is shown in Fig.~\ref{fig:1}. Maser spots were detected within an LSR velocity range of $-$10 km s$^{-1}$ to +6 km s$^{-1}$ with 
the most persistent emission appearing in the $-$2.3 to +1 km s$^{-1}$ range.
If maser emission was seen to persist continuously over three consecutive velocity channels, and also occupied the same region of space within 1$\times$1 mas, those maser spots were considered to be one `maser feature'. Using this criterion we identified 8 maser features from the 19 maser spots detected in our observations. A summary of detections and velocities of maser features and individual spots is given in Table~\ref{table:2}.
The spatial distribution of detected H$_{2}$O maser features is shown in Fig.~\ref{fig:3} with vectors indicating the velocities and directions of those maser features for which proper motions were determinable (\emph{see section 3.2}).
The averaged LSR velocity of maser features is about -1 km s$^{-1}$, which is similar to the LSR velocity of the driving source, as indicated by thermal molecular line observations \citep{Yang02,Ao04,Sunada07}.

Features 1 and 2 differ more than 4.5 km s$^{-1}$ in radial velocity, and have a separation of over 300 mas ($\sim$800AU) from the other maser features. This separation does not exclude them from being a part of the same star forming region, however, due to their absence for the majority of the observing programme, we will not further discuss their nature in this paper.

\null

\subsection{Proper motion and parallax fitting; the distance to the source}
Astrometric maser motions are caused by a combination of periodical offsets due to annual parallax motion, a common proper motion observed for all maser spots caused by the systemic motion of the source with respect to the observer, and internal proper motions inherent to the source caused by phenomena such as expansions or outflows. By tracking the absolute positions of maser features for a period of over two years we successfully disentangled and evaluated the contributions of parallax and proper motions.

Simultaneous parallax and absolute proper motion fitting was done utilising two methods, which were then compared for consistency. Fitting was initially performed for all masers that were detected in at least 4 epochs by assuming that all maser spots are located at the same distance and each spot moves on the sky with a constant velocity. In this paper we call this approach ``group fitting''. Using group fitting we measured an annual parallax of $\pi = 0.367 \pm 0.037$ mas, corresponding to a trigonometric parallactic distance of
$D = 2.72 ^{+0.31}_{-0.25}$ kpc.

For comparison, simultaneous parallax and proper motion fitting was performed for 5 individual maser spots that satisfied the criterion of having been detectable for a year or longer. The results are shown in column 7 of Table~\ref{table:2}. The trigonometric parallaxes obtained were in the range of $\pi =$ 0.252 to 0.430 mas, which agrees well with the value obtained through group fitting. 
Fitting was also done for maser spots that were detected in at least 3 epochs however their duration of less than a year makes them unreliable for estimating parallax, therefore only the information of their proper motions were estimated, using the parallax information from group fitting.

\onecolumn

\begin{table*}[t]
\hspace*{-0.22cm}
\begin{threeparttable}[c]
\caption{H$_{2}$O maser detections, parallax and proper motions for IRAS20143+3634.\label{table:2}}
\begin{center}
\small
\begin{tabular}{ccccccccc}
\hline
Spot&Feature&$V_{\rm LSR}$&Detected &$\Delta \alpha \cos \delta$&$\Delta \delta$&$\pi$ &$\mu_{\alpha}\cos\delta$&$\mu_{\delta}$ \\ 
ID.&ID.&(km s$^{-1}$)&epochs&(mas)&(mas)&(mas)&(mas yr$^{-1}$)&(mas yr$^{-1}$)\\ \hline
A&1&$-9.47$&*****67*&$-225.13$&$-274.67$&&&\\ 
B&2&$6.53$&**3*****&$-169.82$&$-250.17$&&&\\ 
C&2&$5.68$&*23*****&$-169.81$&$-250.12$&&&\\ 
D&2&$6.10$&*23*****&$-169.74$&$-250.19$&&&\\
E&3&$-1.47$&***45***&$-29.10$&$16.25$&&&\\ 
F&3&$-1.05$&*2345***&$-28.91$&$16.73$&&$-3.50\pm0.25$&$-2.90\pm0.57$\\ 
G&3&$-0.63$&***45***&$-28.91$&$16.25$&&&\\
H&3&$-0.63$&*23*****&$-27.87$&$15.81$&&&\\ 
I&4&$-1.89$&****5678&$-24.60$&$11.12$&$0.343\pm0.094$&$-3.03\pm0.19$&$-4.78\pm0.46$\\ 
J&5&$-1.05$&*****678&$-22.02$&$16.58$&&$-2.35\pm0.38$&$-5.31\pm0.87$\\
K&5&$-1.47$&*****678&$-21.86$&$16.64$&&$-2.50\pm0.38$&$-5.32\pm0.87$\\
L&5&$-2.31$&****567*&$-21.14$&$15.94$&&$-3.26\pm0.26$&$-4.45\pm0.61$\\ \hline
&Feature 5 average&-1.61&&&&&$-2.70\pm0.23$&$-5.03\pm0.24$\\ \hline
M&6&$-1.47$&*23*****&$-20.75$&$15.36$&&&\\
N&7&$-0.63$&*****67*&$-9.23$&$10.03$&&&\\
O&7&$-0.21$&*****67*&$-8.88$&$10.98$&&&\\
P&8&$1.05$&***45678&$-0.03$&$0.70$&$0.399\pm0.084$&$-2.71\pm0.14$&$-5.16\pm0.34$\\
Q&8&$0.63$&*2345678&$-0.08$&$0.70$&$0.379\pm0.073$&$-2.75\pm0.09$&$-4.49\pm0.21$\\ 
R&8&$-0.21$&**345*78&$-0.08$&$0.66$&$0.252\pm0.178$&$-2.67\pm0.11$&$-4.88\pm0.27$\\
S&8&$0.21$&12345678&$0.00$&$0.00$&$0.430\pm0.074$&$-2.73\pm0.08$&$-4.49\pm0.21$\\ \hline 
&Feature 8 average&0.42&&&&&$-2.72\pm0.01$&$-4.76\pm0.14$\\ \hline \hline
&&&Group fitting&&&$0.367\pm0.037$&&\\ \hline  
&Avg. of all features&$-1.0\pm0.5$&&&&&$-2.99\pm0.16$&$-4.37\pm0.43$\\
\hline
\end{tabular}
\begin{tablenotes}
\item{Column (4)}: Epoch numbers assigned in column(1) of Table 1, while asterisk represents non-detection.
\end{tablenotes}
\end{center}
\end{threeparttable}
\end{table*}

\null

\twocolumn

\noindent Fig.~\ref{fig:4} shows the resulting fits to the data as offsets in right ascension (R.A.) and declination (Decl.) over the course of our observations. The derived parallaxes and proper motions of individual spots are summarised in Table~\ref{table:2} alongside the result obtained from group fitting.

\begin{figure}[h]
\begin{center}
\vspace{+0.5cm}
\hspace{-1.2cm}
\includegraphics[scale=0.36]{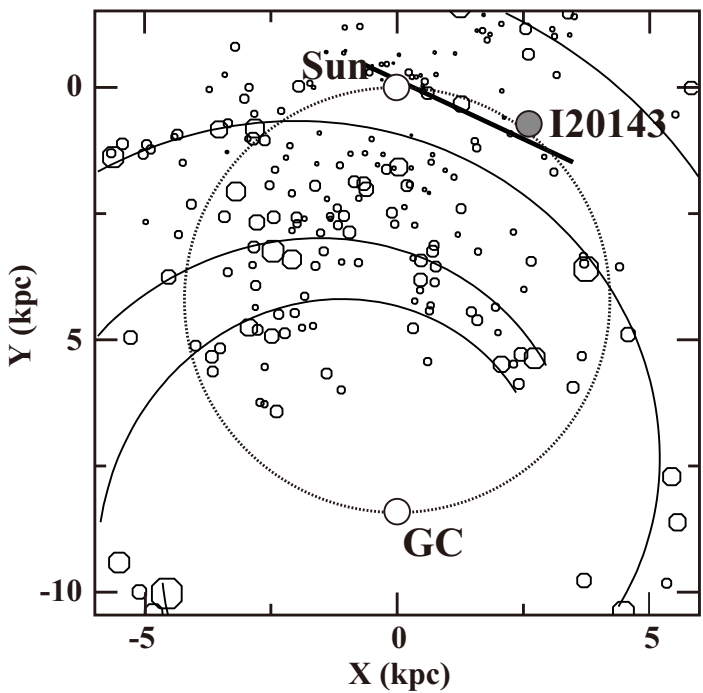}
\caption{The location of IRAS 20143+3634 as seen from a face-on view of the \textbf{Milky Way} (shown as I20143). The dotted circle traces the tangent points. The background is the spiral structure from fig. 5 of \citet{Russeil03} showing the Local arm (solid line), in which they assume $R_0=8.5$ kpc. IRAS 20143+3634 is located near to the tangent point and in the local arm.
\label{fig:2}}
\end{center}
\end{figure}

Our trigonometric parallactic distance is almost a half of the kinematic distance estimates previously reported; 4.4 kpc by \citet{Sunada07} from the peak velocity of NH$_{3}$ lines, and 4.63 kpc by \citet{Yang02} using the velocity of CO lines.  We should note that this large discrepancy is not due to a large source peculiar motion (see section 3.4). It is mainly due to the small velocity gradient against the distance of the Galactic rotation model, which means IRAS 20143+3634 is located where the error of kinetic distance is large.
With Galactic coordinates $(l,b)$=(74$^{\circ}$.57, +0$^{\circ}$.85), our distance estimation indicates that IRAS 20143+3634 resides in the Local arm. It is in the proximity of the tangent point in this direction, based on a Galactocentric distance of the Sun of between 8 and 8.5 kpc (Fig.~\ref{fig:2}, \emph{and see section 4.2}).

\subsection{Internal maser motions}
\begin{figure}[h!]
\begin{center}
\includegraphics[scale=0.85]{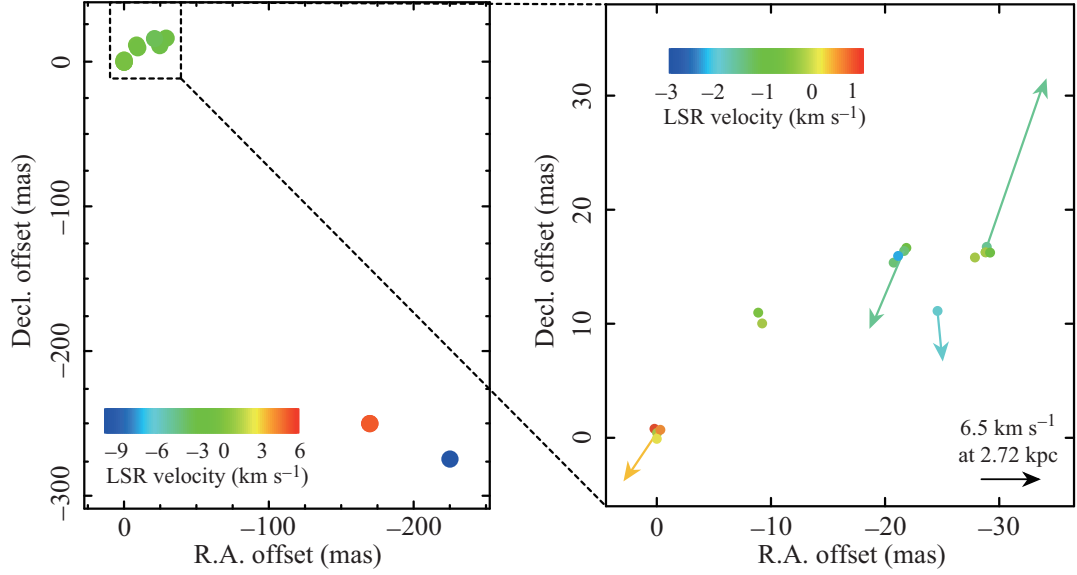}
\caption{Spatial distribution of H$_{2}$O maser spots in IRAS20143+3634.
Colour indicates the LSR velocity of maser spots, and arrows indicate the internal proper motions of maser features.
The origin of the offset is the position of maser spot S with 
$v_{\rm LSR} = 0.21$ km s$^{-1}$ at
$\alpha$ = 20$^{h}$16$^{m}$13$^{s}$.3617, $\delta$ = 36$^{\circ}$43'33.920"(J2000.0)
\label{fig:3}}
\end{center}
\end{figure}

\begin{figure}[h!]
\begin{center}
\includegraphics[scale=0.8]{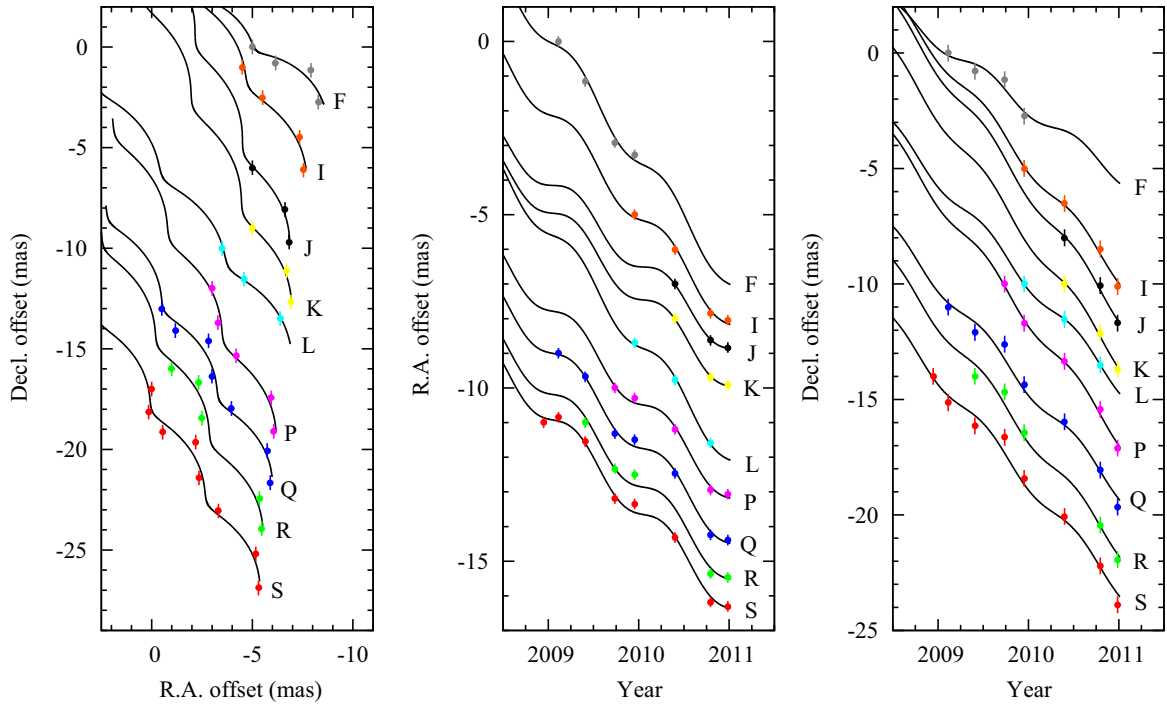}
\caption{
Apparent motions of 9 H$_{2}$O maser spots in IRAS20143+3634 on the sky, with positional errors, and the best-fit curves which were applied to the parallax and proper motion. Each maser spot is represented by an individual colour for clarity.~(\emph{Left}) positions of maser spots in R.A. and Decl. with time running from top-left to bottom-right.~(\emph{middle}) R.A. offsets over time, and~(\emph{right}) Decl. offsets over time.\label{fig:4}}

\end{center}
\end{figure}

The absolute proper motion of maser features is a combination of the source systemic motion and the internal maser motions inherent in the source. By assuming that the internal maser motions are random and thus velocity vectors average to zero we can separate the internal motion from the source motion. This deconstruction was possible for 4 maser features for which the absolute proper motion could be measured. Fig.~\ref{fig:3} illustrates the internal maser motions in IRAS 20143+3634. It can be seen that the more persistent, low-velocity maser features form a group confined to a 16$\times$29 mas ($44 \times 79$ AU using our distance) region. Their transverse velocities were in the range of 2.24 to 24.12 km s$^{-1}$. The nominal dispersion of these velocities is $\sim$10 km s$^{-1}$. 

\null

\subsection{Systemic motion of IRAS 20143+3634}
Estimating the source systemic motion using water masers may be controversial, since the maser emission is not biased to the mass of the source.
In cases where maser emission clearly traces symmetrical structures, such as a bipolar outflow, the dynamic center of the source can be established by tracing the common origin of motion (\emph{see, for example} \citealt{Imai11}). This method assumes that all masers motions are driven by one common driving source. However, in the case where the nature of the driving mechanism is not so obvious it may be misleading to assume that a dynamic center of motion can be accurately found. Previous investigations of water maser sources suggest that the average motion, based on well randomised maser features, gives an accurate estimation of the systemic velocity \citep{Nagayama11a}. In the case of IRAS 20143+3634 the distribution and radial velocities of masers seem randomised, as opposed to showing any bipolar structure. Therefore we estimate the systemic motion of IRAS 20143+3634 using the method demonstrated by \citet{Nagayama11a}.
Using this approach, the vector average of the four maser features for which proper motions were determined was 
($\mu_{\alpha}\cos\delta$, $\mu_{\delta}$) = ($-2.99\pm0.16$, $-4.37\pm0.43$) mas yr$^{-1}$. Where the error value is the standard error of the mean, $\sigma$/$\sqrt{4}$.
Here, $\sigma$ is the standard deviation of the proper motion of 4 features.
This error evaluation was also used in the astrometric observations of WB89-437, ON1, and IRAS05168+3634 (\citealt{Hachi09, Nagayama11a, Sakai12}, respectively).
The use of this method is reasonable for IRAS 20126+3634 
because the averaged LSR velocity of the 4 features is $-1\pm1$ km s$^{-1}$, 
which is consistent with the systemic LSR velocity indicated by the associated 
molecular cloud ($v_{\rm LSR} = -1\pm1$ km s$^{-1}$ \citet{Ao04}).

The observed proper motion is relative to the Sun. To evaluate the motion relative to the LSR, which we call the LSR proper motion, we have to correct for the Solar motion with respect to the LSR. In this paper we use the traditionally defined value of
(U$_{\odot}$,V$_{\odot}$,W$_{\odot}$) = (+10.3, +15.3, +7.7) km s$^{-1}$ (\citealt{Kerr86}, \emph{see also} \citealt{Ando11}).
We calculated the proper motion of IRAS 20143+3634 to be
($\mu_{l}$, $\mu_{b}$)=($-5.75\pm0.33$, $+0.62\pm0.32$) mas yr$^{-1}$ with respect to the LSR, which, using our distance measurement of 2.72 kpc, yields velocities in the Galactic co-ordinates of
($v_{l}$,$v_{b}$)=($-74.1\pm4.3$, $+8.0\pm4.1$) km s$^{-1}$.

\onecolumn

\null

\null

\null

\null

\twocolumn




\section{Discussion}

\subsection{Physical parameters of IRAS 20143+3634}

Mass and luminosity are fundamental parameters of a SFR, whose evaluation depends on the estimated source distance. Our trigonometric distance revises the previous estimate by a factor of about 2. We therefore should revisit past estimates of the physical parameters of IRAS 20143+3634.

\begin{figure}[h!]
\begin{center}
\vspace{+0.5cm}
\includegraphics[scale=0.85]{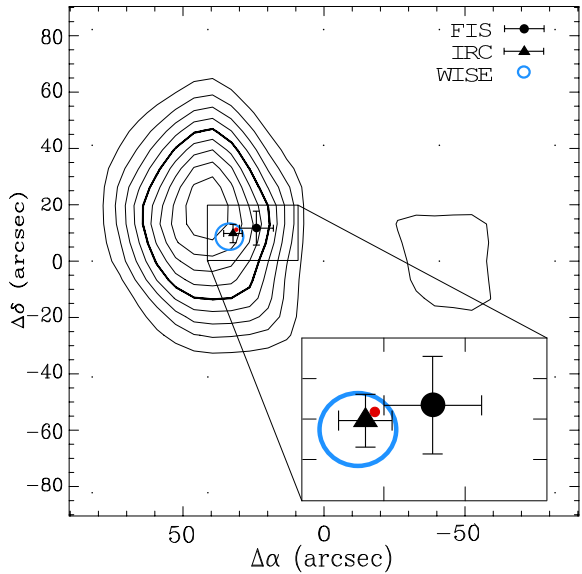}
\caption{A contour map of CS line emission from \citet{Ao04} with positions and accuracies of the sources detected with AKARI-FIS, AKARI-IRC, and WISE \emph{(blue)}. The position of the H$_{2}$O masers is indicated by the small, \emph{(red)} filled circle. The 30"$\times$20" area around the masers is magnified for clarity. \label{fig:5}}
\end{center}
\end{figure}

\cite{Ao04} estimated the core mass of IRAS 20143+3634 to be $M_{\rm LTE} =64 M_{\odot}$ from CS$(J=2-1)$ emission (see Fig.~\ref{fig:5}) under the assumption of local thermodynamic equilibrium (LTE). From velocity dispersion of the same emission they also estimated a virial mass of $M_{\rm vir} =360 M_{\odot}$. With our new distance measurement we re-estimated a core mass of $M_{\rm LTE} = 22 M_{\odot}$ and $M_{\rm vir} =213 M_{\odot}$, and infrared luminosity of $L_\mathrm{IR}=894\,L_\odot$.

The revision of the distance reduces $M_\mathrm{LTE}$ and $L_\mathrm{IR}$, and enlarges the ratio $M_\mathrm{vir} / M_\mathrm{LTE}$. It enforces the trend which \citet{Ao04} show in their fig 14; the less massive cores have a larger excess of $M_\mathrm{vir}$. As discussed in \citet{Ao04}, it  suggests that the turbulent motion in the core dissipates gradually during the formation of the core and subsequent star formation. In their samples IRAS 20143+3634 is in the youngest stage.

\begin{figure}[h!]
\begin{center}
\includegraphics[scale=0.65]{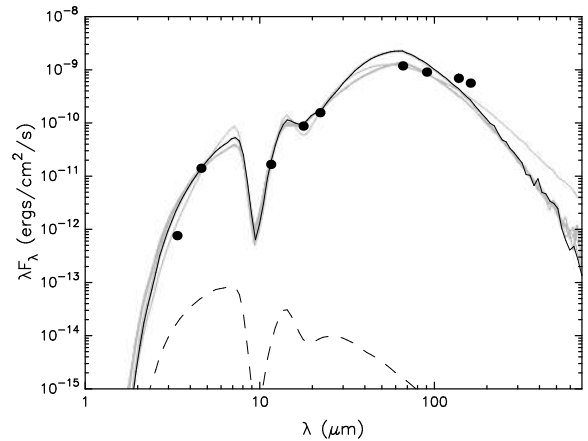}
\caption{SED of MIR and FIR dust emission measured with AKARI at $\lambda=$ 18$\mu$m 65$\mu$m 90$\mu$m 140$\mu$m and 160$\mu$m and WISE at $\lambda=$ 3.4$\mu$m 4.6$\mu$m 12$\mu$m and 22$\mu$m. The thick black line shows the best fit to the photometry data, grey lines represent near-fit models with $\chi^{2}-\chi^{2}_{best} < 3$. The dashed line shows the expected blackbody contribution from an embedded object.\label{fig:6}}
\end{center}
\end{figure}

The spectral energy distribution (SED) is a good indicator of the properties of the energy source. We compiled the SED of IRAS 20143+3634 using photometric data from the AKARI satellite measured with the Infrared Camera (IRC) and Far-infrared Surveyor (FIS) under AKARI point source names 2016133+364333 and 2016128+364336 respectively, and WISE magnitudes under the source name J201613.38+364333.6. These data are employed in the SED shown in Fig.~\ref{fig:6}.

We used a catalogue of young stellar object (YSO) models from \cite{Rob07} and their SED fitting tool to find the best fit to the data via a reduced $\chi^{2}$ approach. The best fit corresponds to a YSO of mass 7.62$M_{\odot}$, stellar temperature of $\sim20,000 K$, luminosity of $3 \times 10^{3}L_{\odot}$ and stellar age of about $2 \times 10^{5}$ years. The best fit model presents an envelope mass of $\sim29 M_{\odot}$ contained within an outer radius of $1 \times 10^5$AU.
Combining our SED model results with re-estimations of the findings of \citet{Ao04} we conclude that IRAS 20143+3634 appears to harbour an intermediate to high mass YSO with outflows, imbedded in a hot molecular core and still accreting material. It is at an early evolutionary stage, resembling a Class I YSO.



\null

\vspace{-0.5cm}

\subsection{The angular velocity of Galactic rotation of the LSR, $\Omega_{0}$.}

\cite{Nagayama11a} demonstrated that the angular velocity of Galactic rotation at the Solar radius, $\Omega_{0}$, which equates to the ratio of the Galactic constants $\Theta_{0} / R_{0}$, can be evaluated using astrometric observations of a source near the tangent point. For a source circularly rotating around the Galactic center on the Galactic plane at the Galactic longitude $l$, it is given by  

\begin{eqnarray}
\Omega_{0} &=& -a_{0}\mu_{l}+v_{r}\left(\frac{1}{D\tan l}-\frac{1}{R_{0}\sin l}\right)\label{eq:3}
\end{eqnarray}

\null

\noindent where $D$ is the distance to the source from the Sun, $v_r$ is its LSR velocity, $\mu_l$ is its systemic proper motion, and $a_0$ is a unit conversion factor from $\mu_l$ to $\Omega_0$ or $a_0=4.74$ km s$^{-1}$ kpc$^{-1}$ (mas yr$^{-1}$)$^{-1}$.
In Equation~(\ref{eq:3}), the second term becomes small for a source located near the tangent point, because $D \tan l=R_0 \sin l$ at the tangent point \citep{Nagayama11a}. Moreover, the LSR velocity of IRAS 20143+3634 is also small. We can, therefore, estimate $\Omega_0$ with improved accuracy from our data.

By applying the observable quantities determined in this paper, namely $D = 2.72 ^{+0.31}_{-0.25}$ kpc, $\mu_{l} = -5.75 \pm 0.33$ mas yr$^{-1}$, and $v_{r} = -1.0 \pm 1.0$ km s$^{-1}$ from the CS$(J=2-1)$ line observations of \citet{Ao04}, we derive an angular velocity of the Galactic rotation at the LSR of $\Omega_{0} = 27.3 \pm 1.6$ km s$^{-1}$ kpc$^{-1}$, when we assume $R_{0} = 8$ kpc.
We demonstrate how the derived value of $\Omega_0$ is less sensitive to the choice value of $R_{0}$ by varying it between the values
$7 \le R_{0} \le 9$ kpc. This only produces a deviation of
$\Delta\Omega_{0} = \pm 0.02$ km s$^{-1}$ kpc$^{-1}$.

Prior to this work, there have been several VLBI astrometric observations of maser sources near the tangent points and the Solar circle. Their estimations of $\Omega_0$ were 
$\Omega_{0} = 28.7 \pm 1.2$ km s$^{-1}$ kpc$^{-1}$ for ON1 \citep{Nagayama11a} and
$\Omega_{0} = 27.3 \pm 0.8$ km s$^{-1}$ kpc$^{-1}$ for ON2N \citep{Ando11}.
In cases where the authors had not done so, we estimated $\Omega_0$ from the data of their observations, using the same method (and the same values in correcting for the Solar motion) as for IRAS 20143+3634. 
$\Omega_{0} = 27.9 \pm 2.6$ km s$^{-1}$ kpc$^{-1}$ for W51 Main/South \citep{Sato10};
$\Omega_{0} = 27.8 \pm 1.0$ km s$^{-1}$ kpc$^{-1}$ estimated using the average of the proper motions of W75N, DR21, DR20 and IRAS20290+4052 in the Cygnus X region \citep{Rygl12};
$\Omega_{0} = 26.3 \pm 1.0$ km s$^{-1}$ kpc$^{-1}$ for W49N \citep{Zhang13}.
Our value of 
$\Omega_{0} = 27.3 \pm 1.6$ km s$^{-1}$ kpc$^{-1}$ for IRAS 20143+3634 agrees very well with these values. With our result included, the mean value of $\Omega_{0}$ estimated this way for tangent point and Solar circle targets is
$\Omega_{0} = 27.6 \pm 0.7$ km s$^{-1}$ kpc$^{-1}$.
All of these results are consistent with each other suggesting that the peculiar motions of these sources are small in the direction of Galactic rotation, and that the sources orbit circularly about the Galactic center. 

Using a multi-parameter fitting procedure \citet{Honma12} estimated $\Omega_{0} = 29.2 \pm 0.8$ km s$^{-1}$ kpc$^{-1}$ for 52 SFRs measured with maser VLBI astrometry.
Also, \citet{Reid04} estimated $\Omega_{0} = 28.2 \pm 0.2$ km s$^{-1}$ kpc$^{-1}$ from the proper motion of Sagittarius A$^{\star}$ using VLBI astrometry.
Furthermore, besides VLBI astrometry, $\Omega_0$ was estimated using the velocity field of stars near the Sun;
$\Omega_{0} = 28.6 \pm 1.4$ km s$^{-1}$ kpc$^{-1}$ for Cephaid variables \citep{Miyamoto98};
$\Omega_{0} = 31.6 \pm 1.4$ km s$^{-1}$ kpc$^{-1}$ for OB stars \citep{Miyamoto98}. The former is consistent with our value and the other VLBI estimations, although most targets of the VLBI estimations are SFR. Finally, since $\Omega_0$ is the ratio of the Galactic constants, we can compare the value calculated using the IAU recommended Galactic constants; $\Omega_{0}=25.9$ km $s^{-1}$ kpc$^{-1}$ from $R_{0}=8.5$ kpc and $\Theta_{0}=220$ km s$^{-1}$. This value is significantly lower than ours and recent estimates.

\begin{figure}[h]
\begin{center}
\vspace{+0.3cm}
\hspace{-0.60cm}
\includegraphics[scale=0.8, angle=0]{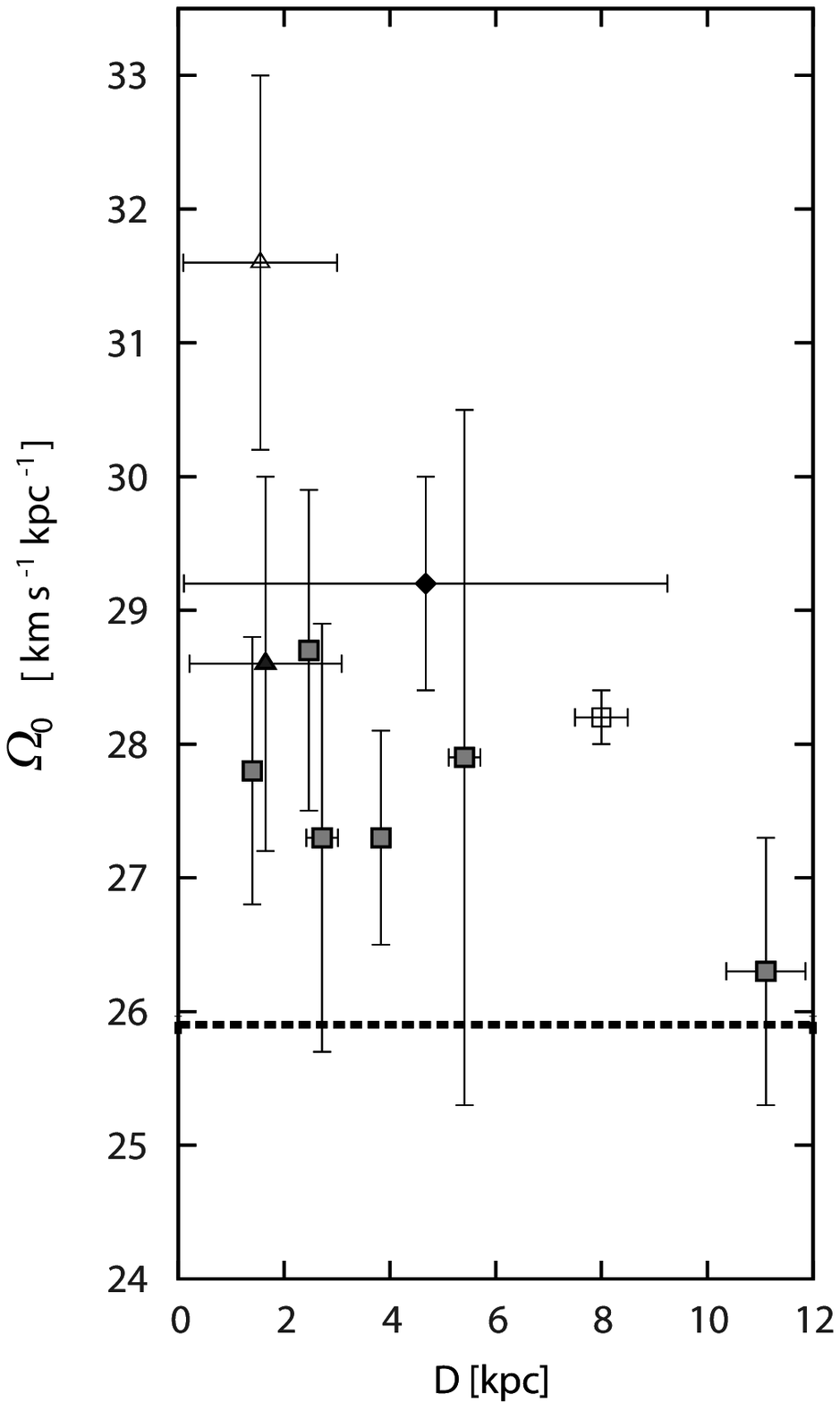}
\caption{Estimated $\Omega_{0}$ using different distance sources. The horizontal bar of each point represents the distance error or range of each sample.
\emph{Filled squares}:~VLBI astrometry of tangent point and Solar circle targets (\emph{In ascending distance: Cygnus X \citep{Rygl12}, ON1 \citep{Nagayama11a}, IRAS 20143+3634 (this paper), ON2N \citep{Ando11}, W51 Main/South \citep{Sato10}, and W49N \citep{Zhang13}).}
\emph{Diamond}:~Multi-parameter fitting for VLBI astrometry of 52 SFRs \citep{Honma12},
\emph{Open square}:~Proper motion of Sgr A* \citep{Reid04},
\emph{Open triangle}:~Velocity field of 1352 OB stars and
\emph{Filled triangle}:~170 Cepheid variables \citep{Miyamoto98},
\emph{Dashed line}:~Calculated using the IAU 1985 recommended values.
\label{fig:7}}
\end{center}
\end{figure}

We have summarised the current estimates of $\Omega_{0}$ graphically, in Fig.~\ref{fig:7}. The horizontal axis denotes the distance to the sources used in each sample. Recent estimates of $\Omega_{0}$ are consistent for a distance range of 1 to 8 kpc
indicating that these estimates are not only valid in the local Galactic neighbourhood but also at larger distances from the Sun. This suggests that these sources rotate axisymmetrically in the disk of the Milky Way with only small peculiar motions.

The value of $\Omega_{0}$ reveals non-linear distortion between the model and the actual velocity field in the disk of the Milky Way. It provides an additional and systematic error for distance estimates that are based on LSR velocities. Therefore, we should use a revised value of $\Omega_0$. Fig.~\ref{fig:7} suggests that the Galactic constants should be set with close attention paid to $\Omega_0$ when they will be revised in the near future.

\subsection{Peculiar motion of IRAS 20143+3634}

To evaluate the peculiar motion of the observed source with respect to the global Galactic motion we must consider the kinematics of the Galaxy. The observational properties of any Galactic sources depend both on the target source motion and that of the Sun. It is difficult to disentangle these two motions. 
With our estimate of $\Omega_0=27.6$ km s$^{-1}$ kpc$^{-1}$, we infer $\Theta_0=221$ km s$^{-1}$ for $R_0=8$ kpc, or $\Theta_0=235$ km s$^{-1}$ for $R_0=8.5$ kpc.
We can fairly assume the rotation velocity of IRAS 20143+3634 to be the same as $\Theta_0$, because its Galactocentric distance is about 0.97$ R_0$; almost the same as the Sun.

Based on these two kinematic models we calculated the peculiar motion of IRAS 20143+3634 to be
$(U_{s},V_{s},W_{s}) = (-0.9 \pm 2.9, -8.5 \pm 1.6, +8.0 \pm 4.3)$ km s$^{-1}$ and
$(U_{s},V_{s},W_{s}) = (-1.0 \pm 2.9, -9.3 \pm 1.5, +8.0 \pm 4.3)$ km s$^{-1}$ for
$R_0=8$ and $R_0=8.5$ kpc, respectively. Here, we again use $v_{r} = -1.0 \pm 1.0$ km s$^{-1}$ from the CS$(J=2-1)$ line observations of \citet{Ao04}.
We adopt the notation used by \citet{Reid09}, where 
$U_{s}$ is the velocity component toward the Galactic center,
$V_{s}$ is the velocity component measured in the direction of Galactic rotation and
$W_{s}$ is the component of motion toward the north Galactic pole.
By giving priority to the value of $\Omega_{0}$, our two models are self-consistent within the errors for values of $8 \le R_{0} \le 8.5$ kpc.
If the Galactic constants were $R_{0} = 8.5$ kpc and $\Theta_{0}=220$ km s$^{-1}$
as given by the IAU recommended values, we instead measure a peculiar motion of
$(U_{s},V_{s},W_{s}) = (3.8 \pm 2.9, -8.5 \pm 1.5,~ 8.0 \pm 4.3)$ km s$^{-1}$.


In all these cases the magnitude of peculiar motion of IRAS 20143+3634 in the Galactic plane is about 10 to 13 km s$^{-1}$, which is consistent with the random velocity of molecular clouds in the galactic disk.
This suggests that the large discrepancy of the previous distance estimation of IRAS 20143+3634, based on its LSR velocity, was not due to the peculiar motion of the source, but was mainly caused by the source being located in a region of the Galaxy where the kinematic velocity falls to a value that approaches the random motion velocity. In such a situation, the kinematic velocity cannot give an appropriate distance and other distance estimation methods should be used.

It is appropriate to mention that the estimated 3-dimensional peculiar motions of SFRs can be influenced by the internal motions of masers, whereby an asymmetrically biased motion traced by water masers in particular could lead to a misinterpretation of the secular motion. A prominent example of an asymmetric outflow is shown in \citet{DeBui06} where masers trace a one-sided jet. Comparison with the radial and proper motions of a core tracing maser species, such as the \emph{class II} methanol maser, is a suitable probe for this effect, highlighting the importance of observing SFRs using multiple maser species.


\section{Conclusions}
By observing H$_{2}$O masers for a period of over two years with VERA, we measured the parallactic distance of IRAS 20143+3634. Our distance of $D = 2.72 ^{+0.31}_{-0.25}$ kpc is 60\% of the previous distance estimate based on the LSR velocity. The new distance  places IRAS 20143+3634 in the Local spiral arm, near to the tangent point in the Cygnus direction.

With our distance estimate, we re-estimated the virial and core masses of IRAS 20143+3634 to be $M_{\rm vir} =213 M_{\odot}$ and $M_{\rm LTE} = 22 M_{\odot}$ using data from \cite{Ao04}, which were 360 $M_\odot$ and 64 $M_\odot$ in the original paper, respectively. The source appears to be more violent than those authors previously thought and has yet to have dissipated turbulent motions inherent in the core. Nevertheless, their conclusions on the general trend of YSOs are still valid. Combining this view with an SED compiled from mid- and far-infrared fluxes from AKARI and WISE, IRAS 20143+3634 appears as a young, 7.62 $M_{\odot}$ intermediate to high mass YSO.

Our distance gives the location of IRAS 20143+3634 to be near the tangent point.
As \citet{Nagayama11a} show, astrometric observations of a source near the tangent point provide the ratio of the Galactic constants $\Omega_{0}$ with little dependence of the Galactocentric distance of the Sun, $R_{0}$. From the astrometry of IRAS 20143+3634 we obtained 
$\Omega_0=27.3\pm1.6$ km s$^{-1}$ kpc$^{-1}$. The value changes only $\pm0.02$  km s$^{-1}$ kpc$^{-1}$, if we vary the value of the Galactocentric distance to the Sun between $7 \le R_{0} \le 9$ kpc.

The values of $\Omega_0$ derived from previous VLBI astrometric observations
of sources near the tangent points and the Solar circle are consistent with each other. This consistency supports the simple circular rotation model of the Milky Way. The average value of these estimates is $\Omega_{0} =27.6\pm0.7$ km s$^{-1}$ kpc$^{-1}$.
This value is also consistent with values based on other procedures and it is worth noting that all but few are higher than that calculated from the ratio of the Galactic constants recommended by the IAU since 1985 \citep{Kerr86}. 
The peculiar motion of IRAS 20143+3634 deviates from the simple circular rotation by about 10 km s$^{-1}$, which is consistent with the random velocity of sources in the Galactic disk.

\null




R.B. would like to acknowledge the Ministry of Education, Culture, Sports, Science and Technology (MEXT), Japan for financial support under the Monbukagakusho scholarship.

\null

T.H. acknowledges support from Strategic Young Researcher Overseas Visits Program for Accelerating Brain Circulation (number R2308) by Japan Society for the Promotion of Science (JSPS).

\null

We would also like to thank the anonymous referee for thoroughly checking our work, and providing the necessary comments and advice that contributed to improving the manuscript.

\bibliographystyle{apj}
\bibliography{arXiv}

\end{document}